\documentclass[structabstract]{aa}  
\usepackage{rotating}
\usepackage{graphicx}
\usepackage{amsmath}
\usepackage[mathscr]{eucal}
\usepackage{natbib}
\bibpunct{(}{)}{;}{a}{}{,} 
\usepackage{txfonts}
\begin{document}
   \title{Detection noise bias and variance in the power spectrum
     and bispectrum in optical interferometry}


	\titlerunning{Bias and variance in the power spectrum and bispectrum}

   \author{J. A. Gordon
          \and D. F. Buscher }

   \institute{Astrophysics Group, Cavendish Laboratory, University of
Cambridge, J.J. Thomson Avenue, Cambridge CB3 0HE, UK
              \email{j.gordon@mrao.cam.ac.uk}}

   \date{Received September xx, xxxx; accepted March xx, xxxx}

  \abstract
  {Long-baseline optical interferometry uses the power spectrum and
    bispectrum constructs as fundamental observables. Noise arising in the
    detection of the fringe pattern results in both variance and biases in the
    power spectrum and bispectrum. Previous work on correcting the biases and
    estimating the variances for these quantities typically includes restrictive
    assumptions about the sampling of the interferogram and/or about the
    relative importance of Poisson and Gaussian noise sources. Until now it has
    been difficult to accurately compensate for systematic biases in data which
    violates these assumptions.}
  {We seek a formalism to allow the construction of bias-free
    estimators of the bispectrum and power spectrum, and to estimate their
    variances, under less restrictive conditions which include both
    unevenly-sampled data and measurements affected by a combination of noise
    sources with Poisson and Gaussian statistics.
  }
  {We used a method based on the moments of the noise distributions to
    derive formulae for the biases introduced to the power spectrum and
    bispectrum when the complex fringe amplitude is derived from an arbitrary
    linear combination of a set of discrete interferogram measurements. 
     }
  {We have derived formulae for bias-free estimators of the power
    spectrum and bispectrum which can be used with any linear estimator of the
    fringe complex amplitude. We have demonstrated the importance of bias-free
    estimators for the case of the detection of faint companions (for example
    exoplanets) using closure phase nulling. We have derived formulae for the
    variance of the power spectrum and have shown how the variance of the
    bispectrum can be calculated.}
  {} 


   \maketitle

\section{Introduction}

Astronomical interferometers combine the light from multiple telescopes to form
interference fringes. The amplitude and phase of the fringes corresponding to
interference between a given pair of telescopes can be combined into a single
complex number, the fringe complex amplitude, which in an ideal interferometer
is proportional to a single Fourier component of the object brightness
distribution.

The complex amplitude forms the fundamental observable in phase-stable
interferometric experiments, for example at radio wavelengths, but at optical
wavelengths the phase of the fringes is unstable on timescales of order
milliseconds, because of optical pathlength perturbations introduced by the
Earth's atmosphere. As a result, observers need to combine the information from
multiple short-exposure interferograms to recover the desired science data.
Since the phase of the fringes is randomly fluctuating, simply averaging the
complex amplitude over multiple interferograms would lead to a vanishingly small
result. Instead two observables, the power spectrum and bispectrum, are
calculated from each interferogram and these can be averaged over frames as both
are immune (to first order) to the atmospheric and instrumental phase
perturbations. The power spectrum serves to capture information about the
modulus of the complex amplitude, while the phase of the bispectrum (also known
as the ``closure phase'') captures information about the phase of the complex
amplitude on closed triplets of baselines.

At the low light levels typically encountered in astronomical interferometry,
the effects of the noise arising in the detection process become important. This
noise can usually be modelled as a mixture of Poisson noise, for example photon
noise, and signal-independent Gaussian noise, for example readout noise in the
detector electronics. We use the term ``detection noise'' to refer to
the combination of all the noise processes in the detection process including
the photon noise.

Detection noise produces random measurement errors which can be reduced by
averaging over many interferograms, but in the case of the power spectrum and
bispectrum it also produces systematic biases in the averaged estimates. There
has been considerable study in the literature on these biases (see for example
\cite{goodman_photon_1976, goodman_precompensation_1976, sibille_infrared_1979,
wirnitzer_bispectral_1985,beletic_deterministic_1989,
perrin_subtractingphoton_2003}). Some authors have
derived formulae for the power spectrum bias which allow for both Poisson and
Gaussian noise sources, but existing bispectrum bias estimates have been derived
under the assumption of Poisson noise only.  As a result, it is difficult to
derive an unbiased estimate of the closure phase from observations where both
Poisson and Gaussian noise are significant. Such conditions are increasingly
present in interferometric observations made at near-infrared wavelengths so the
absence of such an estimator is becoming a critical limitation to the
acquisition of accurate interferometric data as we will demonstrate in Section
\ref{sec:companion}.

Much of the existing power spectrum and bispectrum analysis relies on
the direct Fourier transform of the interferogram to recover the fringe complex
amplitudes. Even under noise-free conditions the use of a Fourier-based
estimator will return inaccurate estimates of the fringe complex amplitude if
the interferogram is, for example, non-uniformly sampled. To account for this,
\citet{tatulli_comparison_2006} introduced a \emph{pixel-to-visibility
transformation matrix} (P2VM) for deriving fringe amplitude estimates in
circumstances where the Fourier-based estimators are inaccurate, which includes
the Fourier estimators as a special case.


There has been considerable study of the variance of the power spectrum
and bispectrum \citep{goodman_photon_1976,goodman_precompensation_1976,
dainty_estimation_1979, sibille_infrared_1979, wirnitzer_bispectral_1985,
roddier_redundant_1987, ayers_knox-thompson_1988}, but only the work of
\citet{tatulli_comparison_2006} applies the more general P2VM approach, and in
this case the variance is calculated under assumptions valid only for purely
Gaussian noise processes. Knowing the variance, and hence the signal-to-noise
ratio is important for predicting the integration times and detection limits
prior to an observation run.


This paper builds on the introduction by \citet{gordon_bias-free_2010} and
previous work by \citet{thorsteinsson_general_2004} to present a general
formalism for computing biases and variances in the presence of detection noise.
Our formalism has two advantages over current treatments: (a) it can be applied
when the fringe complex amplitudes are derived using any linear estimator,
including the P2VM and Fourier transform as special cases and (b) it yields
accurate answers for any additive mixture of Poisson noise and Gaussian noise.
We derive formulae for the biases affecting both the power spectrum and
bispectrum under these conditions and show how our results allow the
construction of \emph{bias-free estimators}. We present simulations to show that
using our new bispectrum estimator can significantly enhance the accuracy of
measuring the closure phase under realistic observation conditions.  We also use
our formalism to derive expressions for the variance of the power spectrum, and
explain the method by which the reader can calculate the bispectrum variance if
required.
\section{The Interferometric Equations}\label{sec:transform}

\subsection{The Interferogram} 
We will start from a specific idealised example and use this to introduce our
more general model which will be applicable to a wide range of realistic
scenarios. In our example, the light from $N_{\rm tel}$ telescopes is combined
to produce a 1-dimensional interference pattern, which consists of the
superposition of a set of sinusoidal fringes at different fringe frequencies
$f^{ij}$ corresponding to the interference between telescopes $i$ and $j$. The
interference pattern is sampled to produce a set of $N_{\rm pix}$ intensity
measurements $\{i_p\}$. We shall call this discrete set of intensity
measurements the ``interferogram''. In our example the interferogram is given
by:
\begin{equation}
  \label{eq:simplesample}
  i_p=\frac{C^{00}}{N_{\rm pix}}+
\frac{1}{N_{\rm pix}}\sum_{i<j}^{N_{\rm tel}}\Re\left[ C^{ij}e^{2\pi{\rm i}\alpha_{p}f^{ij}} \right]
+{\rm noise},
\end{equation}
where $C^{ij}$ denotes the complex amplitude of the fringes corresponding
to interference between light from telescopes $i$ and $j$, $C^{00}$ denotes an offset  due to the
light from all the telescopes (we shall call this the ``zero-spacing flux''),
and $\alpha_p$ is the coordinate of pixel $p$.

Recognising that Eq.~(\ref{eq:simplesample}) represents a sum over a set of
Fourier components, we can derive an estimate $c^{ij}$ of the complex amplitudes
from the sampled intensity data using a discrete Fourier transform (DFT):
\begin{equation}
  \label{eq:simpledft}
c^{ij}=
\sum_{p=1}^{N_{\rm pix}} i_pe^{2\pi{\rm i}\alpha_{p}f^{ij}},
\end{equation}
where we note that this equation can also be used to estimate the zero-spacing
flux by setting $f^{00}=0$.  In noise-free conditions, the estimator
$c^{ij}$ will be equal to the true amplitudes $C_{ij}$ provided that the
samples are evenly spaced in the coordinate $\alpha$ and the frequencies
$f^{ij}$ are integer multiples of the fundamental frequency given by $f_{\rm
  fundamental}=1/[N_{\rm pix}(\alpha_2-\alpha_1)]$. We will call this set of conditions the
``DFT conditions''.

We can rewrite the above equations in the form:
\begin{equation}
  \label{eq:generalforward}
\vec i=\boldsymbol{W}\vec{C}+{\rm noise},
\end{equation}
and:
\begin{equation}
  \label{eq:generalinverse}
\vec c=\boldsymbol{H}\vec{i},
\end{equation}
respectively where $\vec i=[i_1, i_2, \ldots]$ is a vector containing the
pixel intensities, $\vec C=[C^{00}, \mathrm{Re}\{C^{12}\},
\mathrm{Im}\{C^{12}\},\mathrm{Re}\{C^{13}\}, \mathrm{Im}\{C^{13}\}, \ldots]$ is
a vector describing the true complex amplitudes, $\vec c=[c^{00}, \Re\{c^{12}\},
\mathrm{Im}\{c^{12}\},\mathrm{Re}\{c^{13}\}, \mathrm{Im}\{c^{13}\}, \ldots]$ is
a vector describing the estimated complex amplitudes, $\boldsymbol{W}$ is a matrix known as the
``visibility-to-pixel matrix'' (V2PM) and $\boldsymbol{H}$ is the 
pixel-to-visibility matrix (P2VM) mentioned before. When the DFT conditions hold, the
matrices $\boldsymbol{W}$ and $\boldsymbol{H}$ will consist of sine and cosine terms.

If we now assume that the DFT conditions do not hold, for example if the samples
are not evenly spaced or there are ``dead'' pixels, then it is often still
possible to set up a linear set of equations in the form of
Eq.~(\ref{eq:generalforward}) by modifying the matrix $\boldsymbol{W}$. For the
case of any ``working'' beam combiner, we can always find the corresponding matrix $\boldsymbol{H}$ such that:
\begin{equation}
  \label{eq:identity}
\boldsymbol{H}\boldsymbol{W}=\boldsymbol{I}, 
\end{equation}
where $\boldsymbol{I}$ is the identity matrix,
ie.~Eq.~(\ref{eq:identity}) is the condition that $\boldsymbol{H}$ is a
pseudo-inverse of $\boldsymbol{W}$. It is then clear that
Eq.~(\ref{eq:generalinverse}) can be used to derive a value for $\vec c$ such
that $\vec c=\vec C$ under noise-free conditions. The pseudo-inverse
generalises the Fourier inverse to the cases in which the DFT conditions may not
hold; in general the pseudo-inverse matrix will not consist of simple Fourier
terms.

Equation~(\ref{eq:generalinverse}) serves to define the basis of our
more general model for an interferometric measurement. We assume only a discrete set of
intensity measurements $\{i_p\}$ related to a set of complex fringe
amplitudes $\{C^{ij}\}$. We do not require a simple Fourier relationship between
the amplitudes and the pixel intensities; we only require that there exists a
set of estimators for $\{C_{ij}\}$ which are linear in $\{i_p\}$. In other words
we require only that there exists at least one set of complex weights
$\{H^{ij}_p\}$ such that the complex amplitude estimators: 
\begin{equation}
c^{ij}= \sum_{p}i_{p}H^{ij}_{p},
\label{eqn:c_and_H}
\end{equation}
have the property that $c^{ij}=C^{ij}$ in the limit where detection
noise can be neglected. This generalised model can be applied to a
wide variety of beam combination scenarios: the interferograms can be sampled
temporally or spatially or a combination of the two, and multiple telescopes
can be combined ``all-in-one'' into a single multiplexed fringe pattern or
pairwise onto separate detectors.

Note that pseudo-inverse techniques are only presented as one possible
route to deriving $\{H^{ij}_{p}\}$; under circumstances where the forward matrix
$W$ is not known or not fixed (for example if the fringe pattern has not been
spatially filtered) alternative techniques to derive $\{H^{ij}_{p}\}$, perhaps
based on knowing the statistical properties of $W$, may be needed. Nevertheless,
the results presented here will be valid independent of the technique used to
derive values for $H^{ij}_{p}$.

\subsection{Interferometric observables} 
Under phase-unstable conditions, it is common to choose the power spectrum,
$S^{ij}$ and bispectrum, $B^{ijk}$ as the interferometric observables:
\begin{equation}
S^{ij} = \left|C^{ij}\right|^{2},\label{eqn:S_first}
\end{equation}
and:
\begin{equation}
B^{ijk} = C^{ij}C^{jk}C^{ki}.\label{eqn:B_first}
\end{equation}
Typically the values of $C^{ij}$ will change randomly from interferogram to
interferogram due to atmospheric disturbances, but the values of $S^{ij}$ and
$B^{ijk}$ will be more stable. However, unless otherwise specified, we
do not assume any particular distribution for the values of $C_{ij}$; for
example the results will apply equally well under incoherent conditions where
the average value of $C_{ij}$ is zero or conditions where a fringe tracker is
being used and the value of $C_{ij}$ is stable from frame to frame. Our aim is
to determine suitable estimators, $S_{0}^{ij}$, $B_{0}^{ijk}$, for the power
spectrum and bispectrum respectively that do not suffer from bias in the
presence of noise. For this analysis these observables are calculated on an
interferogram-by-interferogram basis and then averaged. Derived quantities, such
as the closure phase, are calculated using the averaged observables, rather than
on an interferogram-by-interferogram basis themselves.

%

%
%

%
\section{Noise Model}
%

The signal recorded by a real detector system will always have some noise
component, which becomes especially important at low light levels. 
We model two distinct sources of noise, photon
shot noise and detector read noise.

\subsection{Photon Noise} 

Photon noise is always present in a real interferogram as a result of the
quantum nature of the detection process which converts the incoming photons to
an electrical signal. For a given pixel $p$, the probability of $N_{p}$
photoevents occurring is given by the Poisson distribution:
\begin{equation}
P_{\textrm{Poisson}}\left(N_{p}|\Lambda_{p}\right) =
\sum_{n}^{\infty}\delta\left(N_{p}-n\right)\frac{\Lambda_{p}^{N_{p}}}{N_{p}!}\exp\left[-\Lambda_{p}\right].\label{eqn:poisson}
\end{equation}%
where $\Lambda_p$ is the ideal noise-free intensity, also called the
``classical intensity'' or ``high-light-level intensity''. Note that this
equation implicitly defines the normalisation of $\Lambda_p$ in units of
photoevents, and we require that $i_p=\Lambda_p$ under noise-free conditions,
thereby defining the normalisation of $i_p$.

\subsection{Read Noise} 

Read noise  is present in all optical and
IR detectors and is often significant at low light levels. Assuming the read
noise is independent of the number of photoevents, the probability of
a given measurement error $\epsilon_{p} = i_{p} - N_{p}$ is given
by the Gaussian distribution:
\begin{equation}
P_{\textrm{Gaussian}}\left(\epsilon_{p}|\sigma_{p}, N_{p}\right) =
\frac{1}{\sqrt{2\pi\sigma_{p}^{2}}}\exp\left[-\frac{\left(\epsilon_{p}\right)^{2}}{2\sigma_{p}^{2}}\right].
\end{equation}
where $\sigma_{p}^2$ is the variance of the noise on pixel $p$; note that the
variance is not assumed to be the same on all pixels.

\subsection{Other Sources of Noise} 

There are additional noise sources that follow
the same statistics as photon or read noise. Detector dark current, $D_{p}$, for
example is also described by the Poisson distribution and could be included in our formalism by
replacing $\Lambda_{p}$ with $\Lambda_{p}+D_{p}$ in Eq.~(\ref{eqn:poisson}). For the sake of readability dark current will not be explicitly included in
this paper.

\subsection{Combined Noise Model} 

If the photon noise and read noise are statistically independent then the
probability of obtaining the noisy data, $i_{p}$, from the underlying ideal
data, $\Lambda_{p}$, is given by:
\begin{equation}
P\left(i_{p}|\Lambda_{p},\sigma_{p}\right) =
\int_{0}^{i_{p}}P_{\textrm{Poisson}}\left(N_{p}|\Lambda_{p}\right)P_{\textrm{Gaussian}}\left(\epsilon_{p}|\sigma_{p},
N_{p}\right)dN_{p}.
\end{equation}
The combined noise distribution is a convolution of the 
Poisson noise and Gaussian noise distributions, so it is convenient to  
derive the moments of our noise model via the moment-generating function of the
probability distribution. The moment-generating function for the combined noise model probability density
$P\left(i_{p}|\Lambda_{p},\sigma_{p}\right)$ is proportional to the product of the
moment-generating functions $M_{\textrm{Poisson}}$ and $M_{\textrm{Gaussian}}$
of the individual distributions:
\begin{equation}
M\left(\nu\right) = M_{\textrm{Poisson}}\left(\nu\right)M_{\textrm{Gaussian}}\left(\nu\right) =
\exp\left[\frac{\sigma_{p}^{2}\nu^{2}}{2}\right]\exp\left[-\Lambda_{p}+\Lambda_{p}
\mathrm{e}^{\nu}\right].
\end{equation}
The first four moments of the
noise model as required to derive the results given in this paper are:
\begin{eqnarray}
\langle i_{p} \rangle = \frac{d}{d\nu}M\left(\nu \right)\Big\lvert_{\nu=0} &=& \Lambda_{p},\label{eqn:noise_model_1} \\
\langle i_{p}^{2} \rangle =\frac{d^2}{d\nu^2}M\left(\nu \right)\Big\lvert_{\nu=0} &=& \Lambda_{p}^{2} + \Lambda_{p} +\sigma_{p}^{2},\label{eqn:noise_model_2} \\ 
\langle i_{p}^{3} \rangle = \frac{d^3}{d\nu^3}M\left(\nu\right)\Big\lvert_{\nu=0} &=& \Lambda_{p}^{3} +3\Lambda_{p}^{2} + \Lambda_{p} +3\Lambda_{p}\sigma_{p}^{2},\label{eqn:noise_model_3}\\
\langle i_{p}^{4} \rangle = \frac{d^4}{d\nu^4}M\left(\nu
 \right)\Big\lvert_{\nu=0} &=& \Lambda_{p}^{4} + 6\Lambda_{p}^{3} +
 7\Lambda_{p}^{2} + \Lambda_{p} + 6\Lambda_{p}^{2}\sigma_{p}^{2} \nonumber\\
 &
 +&6\Lambda_{p}\sigma_{p}^{2} +
 3\sigma_{p}^{4},\label{eqn:noise_model_4}
\end{eqnarray}
where $\langle\ldots\rangle$ denotes taking the expectation over the probability
distribution of the detection noise. For simplicity, only the statistics
of the measured data for fixed
$\{\Lambda_p\}$ is considered at first; averaging over multiple interferograms
will be considered later. As such the angle brackets do \emph{not}
denote taking an average over the distribution of possible values of $\{\Lambda_{p}\}$, due to,
for example, atmospheric distortions to the interferograms.

An important assumption of the following analysis is that the detection noise on different pixels is statistically independent so that for example $\left\langle i_{p1}i_{p2}\right\rangle=\left\langle
i_{p1}\right\rangle\left\langle i_{p2}\right\rangle$ for $p1\neq p2$. It
should be noted that this
assumption will not be true in all cases of practical interest. For example
when temporally-sampling an interferogram with a near-IR detector, it is
common not to reset the charge accumulation capacitors at each pixel between successive time
samples, so that a flux
measurement $i_p$ is derived from the difference between the voltage readings on two
successive samples; in this case the final reading for one sample represents
the initial reading for the next sample and so the noise on the two samples is
correlated. Treating such readout schemes is beyond the scope of the current
paper.

%
\section{Bias-free power spectrum estimator}
%
We seek an estimator $S_0^{ij}$ for the noise-free power spectrum $S^{ij}$ that does not suffer from statistical
bias introduced by the noise on the interferogram. 
A bias-free estimator was calculated by
\citet{tatulli_amber_2007} in the presence of photon noise and read noise, but
we present an alternative derivation here to demonstrate our analysis method. 
For the purposes of evaluating the power spectrum, we drop all
telescope-pair-dependent superscripts, for example writing $H^{ij}_p$ as $H_p$. 

We
start by evaluating the expectation value of the biased estimator
$\left|c\right|^{2}$. Combining Eqs.~(\ref{eqn:c_and_H}) and (\ref{eqn:S_first})
we get:
\begin{equation}
\left\langle
\left|c\right|^{2}\right\rangle=\left\langle\sum_{p1}\sum_{p2}i_{p1}i_{p2}H_{p1}H^{*}_{p2}\right\rangle.\label{eqn:powerspec}
\end{equation}
We split the sum in Eq.~(\ref{eqn:powerspec}) for the cases where
$p1=p2\left(\equiv p\right)$ and for $p1\neq p2$ which gives:
\begin{equation}
 \left\langle \left|c\right|^{2}\right\rangle
 = \sum_{p}\left\langle i_{p}^{2}\right\rangle\left|H_{p}\right|^{2} +
 \underset{\ p1\ \neq\
 p2}{\sum\sum}\left\langle
 i_{p1}\right\rangle\left\langle
 i_{p2}\right\rangle H_{p1}H^{*}_{p2},
\end{equation}
noting that $\left\langle i_{p1}i_{p2}\right\rangle=\left\langle
i_{p1}\right\rangle\left\langle i_{p2}\right\rangle$ for $p1\neq p2$.
Substitution of the moments of the noise model given in
Eqs.~(\ref{eqn:noise_model_1}) and (\ref{eqn:noise_model_2}) leads to:
\begin{equation}
 \left\langle \left|c\right|^{2}\right\rangle= \sum_{p} \left(
 \Lambda_{p}^{2} + \Lambda_{p} + \sigma_{p}^{2}\right) \left|H_{p}\right|^{2} + \underset{\ p1\
 \neq\ p2}{\sum\sum}\Lambda_{p1}\Lambda_{p2}
 H_{p1}H^{*}_{p2}.\label{eqn:powerspec_split}
\end{equation}

The noise-free power spectrum, $S$, can be expressed in terms of the noise-free
interferogram, $\Lambda_p$, as:
\begin{equation}
S = \left|C\right|^{2} =\sum_{p1}\sum_{p2}
\Lambda_{p1}\Lambda_{p2}H_{p1}H^{*}_{p2},\label{eqn:ideal_powerspec}
\end{equation}
This can be split into terms equivalent to those in  
Eq.~(\ref{eqn:powerspec_split}):
\begin{equation}
 S = \sum_{p}  \Lambda_{p}^{2}  \left|H_{p}\right|^{2}+\underset{\ p1\ \neq\  
 p2}{\sum\sum}\Lambda_{p1}\Lambda_{p2}
 H_{p1}H^{*}_{p2},\label{eqn:ideal_powerspec_split}
\end{equation}
which can be substituted into Eq.~(\ref{eqn:powerspec_split}) to give:
\begin{equation}
 \left\langle \left|c\right|^{2}\right\rangle= S + \sum_{p}
 \Lambda_{p}\left|H_{p}\right|^{2},
\end{equation}
and written in terms of the noisy interferograms we see:
\begin{equation}
S = \left\langle \left|c\right|^{2}
\right\rangle -
 \left\langle\sum_{p} \left( i_{p} +
\sigma_{p}^{2}\right) \left|H_{p}\right|^{2}\right\rangle.
\end{equation}
The bias-free power spectrum estimator,
$S_{0}$, is then:
\begin{equation}
S_{0} =  \left|c\right|^{2}
 -\sum_{p} \left( i_{p} +
\sigma_{p}^{2}\right) \left|H_{p}\right|^{2}.\label{eqn:power_spec_est}
\end{equation}

It needs to be remembered that $S_{0}$ is intended to be evaluated on a
per-interferogram, ie.~frame-by-frame basis as atmospheric turbulence
changes the shape of the interferogram between frames. It is straightforward to
show that when this estimator is averaged over a number of frames that the
average is itself unbiased. This is because taking the average over frames
commutes with taking the expectation over the detection statistics, so that:
\begin{equation}
  \label{eq:1}
\left\langle  \frac1{N_{f}}\sum_{f}{S_0}\right\rangle=
 \frac1{N_{f}}\sum_{f}\left\langle{S_0}\right\rangle=
 \frac1{N_{f}}\sum_{f}S
\end{equation}
where $N_{f}$ denotes the number of frames over which the average is taken.
This result depends on the assumption that the detection noise is dependent
only on the value of the flux $\Lambda_p$ in each pixel and is 
statistically independent of the particular shape of the interferogram and
hence of the atmospheric distortions affecting a particular interferogram.
 
For the case where read noise is negligible ($\sigma_{p}=0\mbox{e}^{-}$ and
$\boldsymbol{H}$ is the DFT ($\left|H_{p}\right|^{2} = \left|\mathrm{e}^{2\pi{\rm i}\alpha_{p}f}\right|^2=1$)
Eq.~(\ref{eqn:power_spec_est}) reduces to the estimator reported by \citet{goodman_photon_1976}:
\begin{equation}
S_{1} = \left|c\right|^{2} - N,
\end{equation}
where $N=\sum_{p}i_{p}$, is the total number of photons in an
interferogram. For
high light levels (negligible photon noise and read noise), the
$\left|c\right|^{2}$ term dominates (as this goes as $N^{2}$), and we reduce our
estimator further to the uncorrected estimator (correct only under zero noise conditions):
\begin{equation}
S_{2} = \left|c\right|^{2}.
\end{equation}
%

%
\section{Bias-free power spectrum variance}
%

The derivation for the power spectrum variance requires the first four
moments of the noise model, Eqs.~(\ref{eqn:noise_model_1}) to
(\ref{eqn:noise_model_4}), and stretches to multiple pages due to the $|c^{ij}|^{4}$ term. The derivation
follows the methodology presented above and in the derivation of the bias-free bispectrum
estimator given in Appendix~\ref{app:bispec_der} but is not presented in this
paper. For readability, the formulation of the power spectrum variance,
$\mathrm{var}\left(S_{0}\right)$, is split into terms that dominate in the case
where the noise is dominated by photon noise, read noise or where both noise
sources are significant:
\begin{eqnarray}
&&\mathrm{var}\left(S_{0}\right) = \left\langle\left( \left|c\right|^{2}
 -\sum_{p} \left( i_{p} +
\sigma_{p}^{2}\right) \left|H_{p}\right|^{2}\right)^{2} \right\rangle -
\left( S \right)^{2}\nonumber\\
[1em]& &\mbox{\textbf{Photon Noise Terms:}}\nonumber\\[1em]
&=&2\left|C\right|^{2}\sum_{p}\Lambda_{p}\left|H_{p}\right|^{2}+
CC\sum_{p}\Lambda_{p} H_{p}^{*}H_{p}^{*} + C^{*}C^{*}\sum_{p}\Lambda_{p} H_{p}H_{p}\nonumber\\ 
&+& \left[\sum_{p}\Lambda_{p}\left|H_{p}\right|^{2}\right]^{2} + \sum_{p}\Lambda_{p}
H_{p}H_{p}\sum_{p}\Lambda_{p} H_{p}^{*}H_{p}^{*}\nonumber\\
[1em]& & \mbox{\textbf{Read Noise Terms:}}\nonumber\\[1em]
&+& \sum_{p}\sigma_{p}^{2}\left|H_{p}\right|^{4}+ \left[ \sum_{p}\sigma_{p}^{2} \left|H_{p}\right|^{2} \right]^{2}+ \sum_{p}\sigma_{p}^{2}H_{p}H_{p}\sum_{p}\sigma_{p}^{2}H_{p}^{*}H_{p}^{*}\nonumber\\
[1em]& & \mbox{\textbf{Coupled Terms:}}\nonumber\\[1em]
&+& 2 \sum_{p}\Lambda_{p}
\left|H_{p}\right|^{2}\sum_{p}\sigma_{p}^{2}\left|H_{p}\right|^{2} +\sum_{p}\Lambda_{p}
H_{p}H_{p}\sum_{p}\sigma_{p}^{2}H_{p}^{*}H_{p}^{*}\nonumber\\
&+&\sum_{p}\Lambda_{p}H_{p}^{*}H_{p}^{*}\sum_{p}\sigma_{p}^{2}H_{p}H_{p}
+ 2\left|C\right|^{2}\sum_{p}\sigma_{p}^{2}\left|H_{p}\right|^{2}\nonumber\\
&+&CC\sum_{p}\sigma_{p}^{2}H_{p}^{*}H_{p}^{*}+
C^{*}C^{*}\sum_{p}\sigma_{p}^{2}H_{p}H_{p}- 2
C\sum_{p}\sigma_{p}^{2}\left|H_{p}\right|^{2}H_{p}^{*}\nonumber\\ 
&-& 2C^{*}\sum_{p}\sigma_{p}^{2}\left|H_{p}\right|^{2}H_{p}.\nonumber
\label{eqn:powspecvar}
\end{eqnarray}
For comparison with previous work we rewrite Eq.~(\ref{eqn:powspecvar}) assuming
DFT conditions, that read noise is constant across all pixels
($\sum_{p}\sigma^{2}_{p}=N_{\mathrm{pix}}\sigma^{2}$), and that atmospheric
phase fluctuations are random and as such the cross-product of independent
complex quantities average to zero:
\begin{eqnarray}
\mathrm{var}\left(S_{\textrm{DFT}}\right)&=&\overbrace{2N\left|C\right|^{2} + 
N^{2} +
\sum_{p}\Lambda_{p} \mathrm{e}^{4\pi{\rm i}\alpha_{p}f^{ij}}\sum_{p}\Lambda_{p}
\mathrm{e}^{-4\pi{\rm i}\alpha_{p}f^{ij}}}^{\mathrm{Photon\ Noise\
Terms}}\nonumber\\ &+& \overbrace{
 N_{\mathrm{pix}}\sigma^{2}+ N_{\mathrm{pix}}^{2}\sigma^{4}}^{\mathrm{Read\
Noise\ Terms}}\nonumber\\ [1em]
&+& \overbrace{2\left|C\right|^{2}N_{\mathrm{pix}}\sigma^{2} +
2NN_{\mathrm{pix}}\sigma^{2}}^{\mathrm{Coupled\ Terms}}.
\label{eqn:powspecvar2}
\end{eqnarray}

\begin{figure}[tb]\centering
\includegraphics[width=1.0\columnwidth]{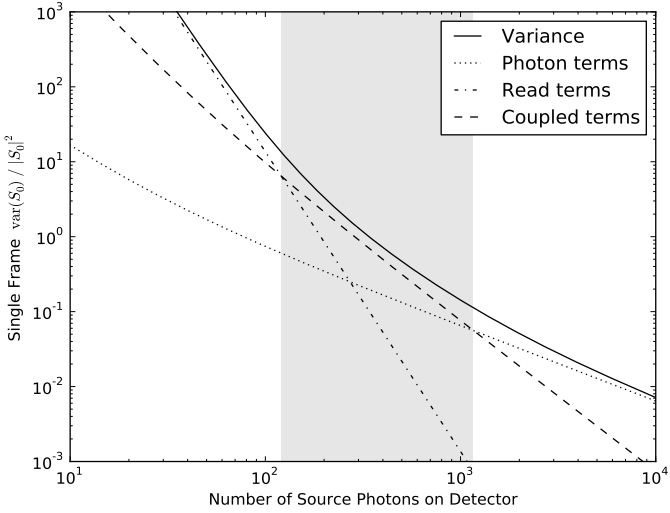}
\caption{The relative importance of the photon, read
($\sigma_{p}=3\mbox{e}^{-}$) and coupled noise terms under DFT conditions for
the power spectrum variance with respect to the number of source photons
detected. The area highlighted in grey is the region where the coupled terms
dominate the power spectrum variance.}
\label{fig:pow_spec_var}
\end{figure}

Separate analysis of the photon and detector noise terms, as is often done (see
e.g. \citet{perrin_squared_2005, le_bouquin_imagerie_2005}), means that the
Gaussian-Poisson coupled terms are missed. Inspection of
Eq.~(\ref{eqn:powspecvar2}) and Fig.~\ref{fig:pow_spec_var} show the
importance of including these coupled terms for a wide range of photon fluxes,
and therefore of incorporating both the photon noise and read noise into a
single noise model. In the above example, for $\sim300$ source photons
on the detector, the variance on the powerspectrum is underestimated by a factor
of over 6 if cross-terms are excluded from the variance calculation.
%
\section{Bias-free bispectrum estimator}
%

The bias-free bispectrum estimator is calculated in the same manner as the
bias-free power spectrum estimator. The definition of the bispectrum presented
in Eq.~(\ref{eqn:B_first}) expands to:
\begin{equation}
B^{ijk} = C^{ij}C^{jk}C^{ki} =
\sum_{p1}\sum_{p2}\sum_{p3} \Lambda_{p1}\Lambda_{p2}\Lambda_{p3}
H^{ij}_{p1}H^{jk}_{p2}H^{ki}_{p3}.\label{eqn:bispec_expanded}
\end{equation}
Splitting the sum in Eq.~(\ref{eqn:bispec_expanded}) gives $5$
terms rather than $2$ as for the power spectrum estimator. As such the
derivation is quite lengthy and is given in Appendix~\ref{app:bispec_der}. We present here the
bias-free bispectrum estimator, $B^{ijk}_{0}$:
\begin{eqnarray}
B^{ijk}_{0}&=& 
c^{ij}c^{jk}c^{ki}\nonumber\\
 &-&  c^{ij}\sum_{p}\left(i_{p} +
 \sigma^{2}_{p}\right) H^{jk}_{p}H^{ki}_{p} \nonumber\\
 &-&\ c^{jk}\sum_{p}\left(i_{p} +
 \sigma^{2}_{p}\right) H^{ij}_{p}H^{ki}_{p} \nonumber\\ 
 &-& c^{ki}\sum_{p}\left(i_{p} +
 \sigma^{2}_{p}\right) H^{ij}_{p}H^{jk}_{p} \nonumber\\
 &+& \sum_{p}\left(2 i_{p}
 +3\sigma^{2}_{p}\right)H^{ij}_{p}H^{jk}_{p}H^{ki}_{p}.
 \label{eqn:bias_free_bispec}
\end{eqnarray}
This equation has been verified though computational simulations to the
0.1\% level. In sections \ref{sec:ccp_bias} and \ref{sec:companion} we present
the results of these simulations regarding the closure phase, derived from the
bias-free bispectrum estimator in Eq.~(\ref{eqn:bias_free_bispec}).

If we assume DFT conditions so that 
$H_p^{ij}=\mathrm{e}^{2\pi{\rm i}\alpha_{p}f^{ij}}$ and
and an all-in-one interference pattern such that 
$f^{ij}+f^{jk}+f^{ki}=0$, then we have
$H^{jk}_{p}H^{ki}_{p}=H_{p}^{ji}$ where we note $H_{p}^{ji} =
\left(H_{p}^{ij}\right)^{*}$ and $H^{ij}_{p}H^{jk}_{p}H^{ki}_{p}=1$. 

If we also assume the read noise $\sigma_{p} = \sigma$ is constant
across all pixels rather than varying on a pixel by pixel basis then:
\begin{equation}
\sum_{p}\sigma^{2}\mathrm{e}^{2\pi{\rm i}\alpha_{p}f^{ji}}=0. 
\label{eqn:const_sigma_dft}
\end{equation}
The fact that the third order moment of the Gaussian distribution is
zero would suggest that there should be no read noise terms under DFT conditions
for constant $\sigma$, however it must be remembered that there are second order
moments in the bias correction terms which will contribute to the bias under
these conditions.

If we
further assume that read noise is negligible ($\sigma_{p}=0$), then $B^{ijk}_{0}$ reduces to the estimator
presented by \citet{wirnitzer_bispectral_1985} accounting for the bias introduced by photon
noise alone:
\begin{eqnarray}
B^{ijk}_{1} &=&
 c^{ij}c^{jk}c^{ki}  \nonumber\\
&-&\left|c^{ij}\right|^{2}
-\left|c^{jk}\right|^{2}
-\left|c^{ki}\right|^{2}
+2 N,\label{eqn:hist_biaspec}
\end{eqnarray}
where $N=\sum_{p}i_{p}$ is the mean number of photons in the interferogram.

For high light levels (negligible photon noise and read noise), the
$c^{ij}c^{jk}c^{ki}$ term dominates (as this goes as $N^{3}$), and we reduce our
estimator further to the uncorrected estimator (correct only under zero noise conditions):
\begin{equation}
B^{ijk}_{2} = c^{ij}c^{jk}c^{ki}.
\end{equation}
%

%

The bispectrum variance is not presented in this paper as the
$|c^{ij}c^{jk}c^{ki}|^{2}$ term leads to a long and tedious derivation. The
zealous reader is invited to calculate the expression for the bispectrum
variance if required using the method presented for calculating the bispectrum
in Appendix~\ref{app:bispec_der}.

The remainder of this paper will be concerned with demonstrating the magnitude of
the bias introduced when the wrong estimator is used in the presence of photon
and read noise at low light levels.
\section{Closure phase bias}\label{sec:ccp_bias}

\begin{figure}[tb]\centering
\includegraphics[width=1.0\columnwidth]{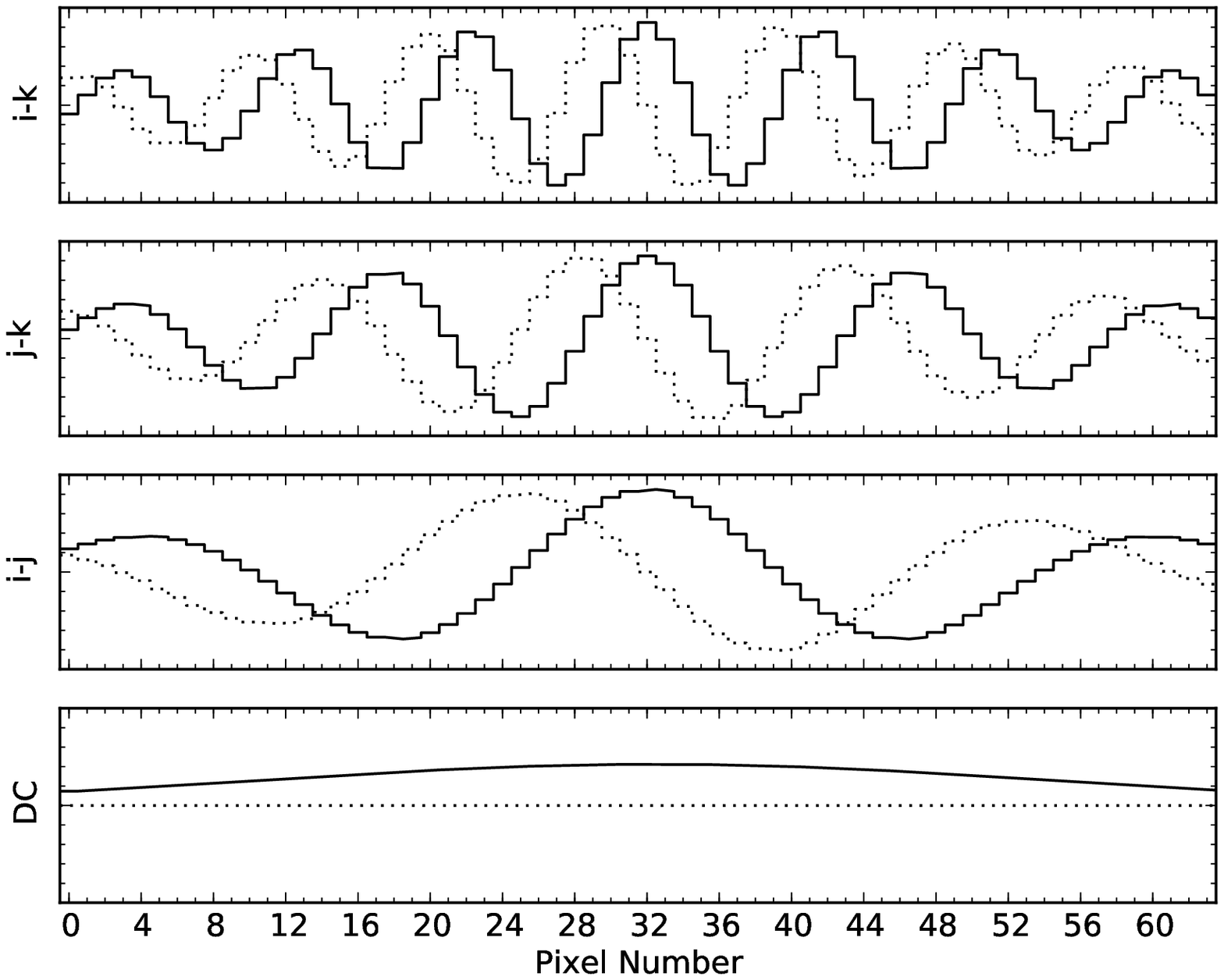} \caption{The transform
matrix, $\boldsymbol{W}$ under non-DFT conditions. The DC term shows the beam
profile adopted, and the cosine (solid lines) and sine (dotted lines) components
of the fringe pattern for each baseline show the deviation from integer number
of fringe periods across the detector (with frequencies in the ratio
$1.1\times\{1:2:3\}$). This is similar (but with the addition of a DC term) to
the visibility to pixel matrix introduced by
\citet{le_bouquin_pupil_2006}.}
\label{fig:v2pm}
\end{figure}

\begin{figure*}[tb]\centering
\includegraphics[width=1.0\textwidth]{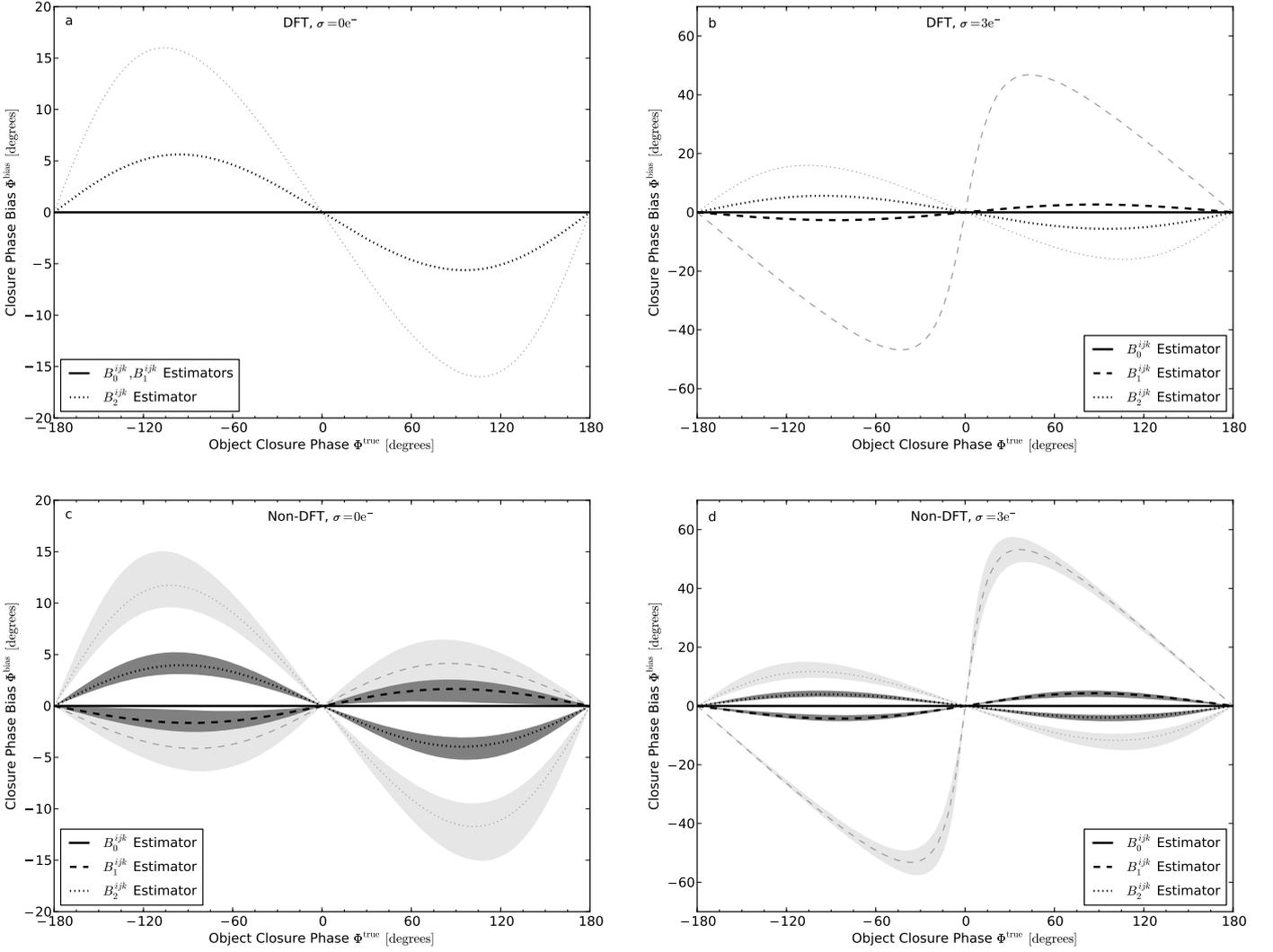}
\caption{The bias in closure phase for $B^{ijk}_{0}$ (solid line),
 		$B^{ijk}_{1}$ (dashed line), and $B^{ijk}_{2}$
 		(dotted line) estimators under a number of measurement conditions. Plots
 		\textbf{a} and \textbf{b} show DFT conditions, whilst plots
 		\textbf{c} and \textbf{d} show non-DFT conditions (described in the text).
 		Read noise of $\sigma_{p}=3\mbox{e}^{-}$ is present in plots \textbf{b} and
 		\textbf{d}. The thin and think lines show 120 and 300 photons
 		reaching the detector respectively. The shaded regions in plots
 		\textbf{c} and \textbf{d} show the range of possible biases resulting from
 		different possible combinations of phases on the three baselines contributing
 		to the source closure phase.}
 \label{fig:closure_phase_bias}
\end{figure*}

To illustrate the magnitude of the systematic bias in a realistic observation we
present the closure phase data from a set of simple simulations. We created
ideal interferograms resulting from a three telescope interferometer observing a
target with a known and controllable closure phase. Photon noise and read noise
were added, and the bispectrum (and hence closure phase) extracted using each of
the three estimators presented above. The interferograms were created
under either DFT or non-DFT conditions. The fringe frequency on the detector for
each baseline was set at a ratio $\{1:2:3\}$ for the DFT case, and
$1.1\times\{1:2:3\}$ for the non-DFT conditions. To simulate non-DFT conditions
we rewrite Eq.~(\ref{eq:simplesample}) as:
\begin{equation}
i_{p} = b_{p}C^{00} + b_{p}\sum_{i < j}^{N_{\rm tel}}
\Re\left[ C^{ij}e^{{\rm i}2\pi \alpha_{p}
f^{ij}}\right]+{\rm noise},\label{eq:simple_more_complex}
\end{equation}
where the beam profile, $b_{p}$ is either flat: $b_{p}=1/N_{\rm pix}$ for DFT
conditions (as in Eq.~(\ref{eq:simplesample})), or a cropped Gaussian of the form:
\begin{equation}
b_{p}=\frac{\exp\left[-\frac{\left(p-\mu\right)^{2}}{2\varsigma^{2}}\right]}
{\sum_{p'}^{N_{\rm
pix}}\exp\left[-\frac{\left(p'-\mu\right)^{2}}{2\varsigma^{2}}\right]},
\end{equation}
where $\varsigma=22$ pixels and $\mu$ is centered
on the $64$ pixel detector. From
Eq.~(\ref{eq:simple_more_complex}) we can construct a V2PM, $\boldsymbol{W}$,
which is correct under DFT and non-DFT conditions. The form of $\boldsymbol{W}$ is shown in Fig.~\ref{fig:v2pm} for the
non-DFT conditions described. We use the Singular Value Decomposition method to
find the optimum (in terms of least squares) pseudo-inverse of $\boldsymbol{W}$, which gives us
$\boldsymbol{H}$. An instrumental visibility loss was introduced such
that the maximum visibility on a baseline is $|V_{i}|=1/3$. The results
presented in Fig.~\ref{fig:closure_phase_bias} are calculated using the ideal
interferograms and therefore show only the closure phase bias and suffer no
variance contribution. These are given by:
\begin{eqnarray}
B^{ijk}_{2} &=& B^{ijk} + \sum_{p}\Lambda_{p}\mathrm{e}^{2\pi{\rm
i}\alpha_{p}f^{ij}}\sum_{p}\left(\Lambda +
\sigma^{2}\right)\mathrm{e}^{2\pi{\rm i}\alpha_{p}f^{ji}}\nonumber\\
&+& \sum_{p}\Lambda_{p}\mathrm{e}^{2\pi{\rm
i}\alpha_{p}f^{jk}}\sum_{p}\left(\Lambda +
\sigma^{2}\right)\mathrm{e}^{2\pi{\rm i}\alpha_{p}f^{kj}}\nonumber\\
&+& \sum_{p}\Lambda_{p}\mathrm{e}^{2\pi{\rm
i}\alpha_{p}f^{ki}}\sum_{p}\left(\Lambda +
\sigma^{2}\right)\mathrm{e}^{2\pi{\rm i}\alpha_{p}f^{ik}} +
N,\label{eq:ideal_cp_2}
\end{eqnarray}
and:
\begin{equation}
B^{ijk}_{1} = B^{ijk}_{2} - S^{ij} - S^{jk} -
S^{ki} - N - 3N_{\rm pix}\sigma^{2},\label{eq:ideal_cp_1}
\end{equation}
whilst the bias free estimator is simply:
\begin{equation}
B^{ijk}_{0} = B^{ijk}.
\end{equation}
We note the order of the indices in the DFT terms in
Eqs.~(\ref{eq:ideal_cp_2}) and (\ref{eq:ideal_cp_1}) and that
the ideal source bispectrum and power spectrum are independent of the
measurement system (weather or not if meets the DFT conditions), whereas
the bias terms implicitly assume DFT conditions for the $B^{ijk}_{1}$ and $B^{ijk}_{2}$
estimators. Fig.~\ref{fig:closure_phase_bias} shows a simulation of the bias
present on the closure phase for the three estimators for simulations of 120 and 300 photons per interferogram reaching the detector.

\subsection{Zero read noise} 

In the situation where photon noise dominates the
noise model ($\sigma_{p}=0\mbox{e}^{-}$) and the instrument is setup such that the
interferograms are formed under DFT conditions, $B^{ijk}_{0}$ and
$B^{ijk}_{1}$ both return the correct closure phase, however
$B^{ijk}_{2}$ is affected by a bias term resulting from photon
noise. At these light levels, the bias in $B^{ijk}_{2}$
becomes significant for the majority of source closure phases.

\subsection{Non-zero read noise} 

At low light levels the read noise on the detector
becomes important and comparable to the photon noise.
Fig.~\ref{fig:closure_phase_bias}b shows simulations under DFT conditions, as
in Fig.~\ref{fig:closure_phase_bias}a, but with $\sigma_{p}=3\mbox{e}^{-}$ read
noise included. This level of read noise is quite low compared to many of the
optical detectors in use today. We see that failing to account for read noise introduces
a significant bias in closure phase estimates for both $B^{ijk}_{1}$ and
$B^{ijk}_{2}$. It is clear that the magnitude of the bias term is
strongly dependent on the source brightness, and that $B^{ijk}_{2}$ is
preferable to $B^{ijk}_{1}$ under certain combinations of noise and photon flux.


\subsection{Non-DFT Conditions} 

Deviation from DFT conditions changes the way statistical bias affects the
bispectrum. Inspection of Eq.~(\ref{eqn:bias_free_bispec}) shows that the
statistical bias can affect both the real and imaginary parts of the bispectrum
when $\boldsymbol{H}$ is not the DFT. If the measurement system conforms
to DFT conditions, and the read noise is negligible or constant across all pixels, the
bias only affects the real part of the bispectrum estimators as shown when
applying Eq.~(\ref{eqn:const_sigma_dft}) to Eq.~(\ref{eqn:bias_free_bispec}).
It is also apparent that the magnitude of the bias is no longer dependent only on the
value of the closure phase, but on the value of the individual phase of each
baseline. The non-uniform beam profile causes an overlap of the fringe
components in frequency space, and the non-integral fringe frequency on the detector
invalidates the assumption that $H^{jk}_{p}H^{ki}_{p}=H_{p}^{ji}$ and
$H^{ij}_{p}H^{jk}_{p}H^{ki}_{p}=1$ which led to the formulation of
$B^{ijk}_{1}$. Fig.~\ref{fig:closure_phase_bias}.c is provided as an example of
the impact of non-DFT conditions on the closure phase bias for a photon noise
dominated instrument ($\sigma_{p}=0$). Note that the non-DFT conditions change
the bias terms, meaning $B^{ijk}_{1}$ is now biased when it was not under DFT
conditions. Fig.~\ref{fig:closure_phase_bias}.d shows the effect of non-DFT
conditions on a system with photon and read noise and again there is an alteration of the
bias terms which $B^{ijk}_{0}$ corrects but the other estimators do not.
The grey regions on the non-DFT plots show the range of biases
associated with each estimator depending on the individual phases of the triplet
producing a given closure phase. The lines within the shaded regions represent
the expectation of the closure phase bias over all possible individual phases.

We see that the bias on the closure phase does not cancel when
accumulating frames; specifically, that there is no advantage to observing in
phase-unstable conditions over phase-stable conditions as a means of reducing
bias. Therefore, bias-wise, one should always favor fringe-tracker assisted
observations whereby the longer discrete integration times result in higher SNR,
even in the photon-rich regime where fringe tracking is not strictly necessary.

\subsection{Closure phase bias and statistical uncertainty}

\begin{figure}[tb]\centering
\includegraphics[width=1.0\columnwidth]{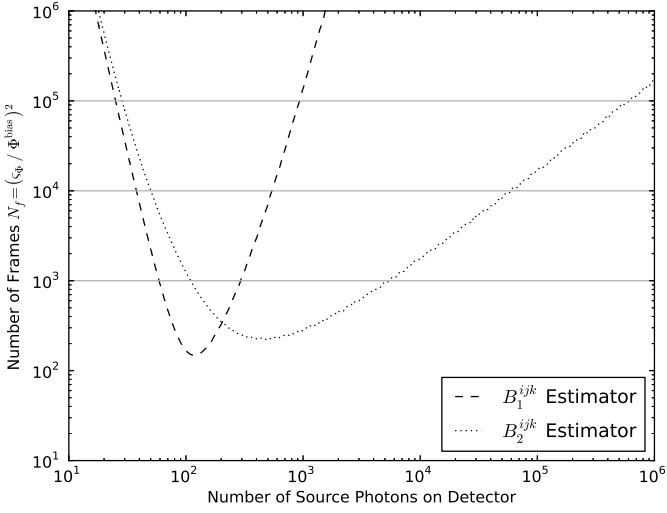}
\caption{The number of integrated frames required to reduce the
statistical uncertainty to the level of the bias for the $B^{ijk}_{1}$ (dashed)
and $B^{ijk}_{2}$ (dotted) estimators.}
\label{fig:closure_phase_var}
\end{figure}

The bias is only significant at low light levels, and tends to zero as
the light level increases. At low light levels the variance is also high, and we
must determine the relative importance of these two processes. We present
Fig.~\ref{fig:closure_phase_var} which shows the relative importance of bias,
$\Phi^{\mathrm{bias}}$, and closure phase standard deviation,
$\varsigma_{\Phi}$, using different estimators.

We show the number of frames, $N_{f}$, such that:
\begin{equation}
\Phi^{\mathrm{bias}} = \frac{\varsigma_{\Phi}}{\sqrt{N_{f}}},
\end{equation}
which is the number of interferograms required to reduce the magnitude
of the statistical uncertainty to that of the bias.
The closure phase standard deviation per frame, $\varsigma_{\Phi}$, is given
by:
\begin{equation}\label{eq:cp_std_dev}
\varsigma_{\Phi} =
\frac{1}{\left|B^{ijk}\right|}\times\sqrt{\mathrm{var}\left[\mathrm{Im}\left(B^{ijk}_{f}\times\frac{B^{ijk*}}{\left|B^{ijk}\right|}\right)\right]},
\end{equation} 
where $B^{ijk}_{f}$ is the bispectrum calculated per frame using the
chosen estimator and
$\mathrm{Im}\left(B^{ijk}_{f}\times\frac{B^{ijk*}}{\left|B^{ijk}\right|}\right)$
gives the projection of $B^{ijk}_{f}$ on a unit vector perpendicular to
$B^{ijk}$, and. This method is used because it is insensitive to wrapping errors
in the complex plane when the variance is large. In this example, the magnitude
of the bias and variance are measured when the source closure phase is at
$90^{\circ}$. The values presented in Fig.~\ref{fig:closure_phase_var} were derived
using numerical simulations to the 0.1\% level, and the bias contribution
was verified analytically using Eqs.~(\ref{eq:ideal_cp_2}) and
(\ref{eq:ideal_cp_1}).

It can be seen in Fig.~\ref{fig:closure_phase_var} that the bias is
important for realistic numbers of frames (e.g.~$10^{3}$ frames or 25s
integration for 25ms exposures) compared to the uncertainty for a wide range of typical light
levels.

\section{Simulation: Detection of faint companions}\label{sec:companion}

\begin{table}[htbp]
\caption{\label{tab:sim_params}Simulation Parameters}
\centering
\begin{tabular}{p{5.7cm}l}
\hline\hline
\noalign{\smallskip}
Parameter&Value\\
\noalign{\smallskip}
\hline
\noalign{\smallskip}
Primary Raduis\dotfill							& $R_{\star}$\\
Primary K-band Magnitude \dotfill	& 5.0 Mag\\
Companion Separation ($s$)\dotfill				& $\left(10, 100, 1000\right)R_{\star}$\\
Companion flux ratio ($r$)\dotfill 				& 1\%\\
\noalign{\smallskip}
\hline
\noalign{\smallskip}
Telescope Diameter\dotfill 		& 1.8 m\\
Number of Apertures\dotfill 	& 3\\
Instrumental Baselines\dotfill	& $\{1:2:3\}$\\
Optical Throughput\dotfill 		& 1\%\\
Instrumental Visibility\dotfill & $1/3$\\
Center Wavelength\dotfill		& $2.211\mu$m\\
Bandwidth\dotfill				& 48 nm\\
Exposure Time\dotfill 			& 25 ms\\
Number of Frames\dotfill		& $10^{3}, 10^{4}, 10^{5}$\\
Total Integration Time\dotfill	& 25s, 4m10s, 41m40s\\
Read Noise\dotfill				& $\sigma=3\mbox{e}^{-}$\\
Quantum Efficiency\dotfill 		& 80\%\\
Science Detector Pixels\dotfill	& 64\\
\noalign{\smallskip}
\hline
\end{tabular}
\tablefoot{The top and bottom panels show the source and instrument
parameters respectively.}
\label{tab:cpn}
\end{table}

Closure phase nulling is an observational techniques that allows the detection
of faint companions \citep{ chelli_phase_2009}.
The technique has been demonstrated by \citet{duvert_phase_2010}, detecting a 5
magnitude fainter companion using AMBER/VLTI around HD 59717.

Closure phase nulling uses the fact that the closure phase jumps from 0 to 180
degrees as the visibility function of the primary in the system passes through a
null. The presence of a companion becomes evident as a deviation from an
instantaneous phase jump. The profile of the phase jump gives information about
the projected separation and flux ratio of the system components. We show that
statistical bias in the closure phase leads directly to incorrect estimation of the system
properties.

We follow \citet{chelli_phase_2009} and simulate a binary system where the
primary is described by a uniform disk of radius, $R_{\star}$, with a point
source companion at a separation, $s$, with flux ratio, $r$. For simplicity we
simulate a three telescope interferometer such that the position angle of the
binary system is parallel to the frequency axis of the interferometer. The
interferometer samples three frequencies $\{{u_{12}, u_{23}}, u_{13}\}$ such
that $u_{13}=u_{12}+u_{23}$. The instrumental parameters are given in
Table \ref{tab:cpn} and we assume the instrument is functioning under DFT conditions.

The complex visibility,
$C^{ij}(u)$, at a given frequency, $u$, is described by:
\begin{equation}
C^{ij}(u) = \frac{V_{\star}(u) + r \mathrm{e}^{{\rm i}2\pi us}}{1+r},\label{eqn:comp_vis_binary}
\end{equation}
which is the visibility function, $V_{\star}(u)$, of the primary:
\begin{equation}
V_{\star}(u) = 2\frac{J_{1}(2\pi u R_{\star})}{2\pi u R_{\star}},
\end{equation}
plus the contribution of the secondary with a phase modulation related to the
projected separation, $s$, and flux ratio, $r$. $J_{1}$ is the first order Bessel function and
$R_{\star}$ is the radius of the primary. 

\begin{figure*}[tb]\centering
   		\includegraphics[width=1.0\textwidth]{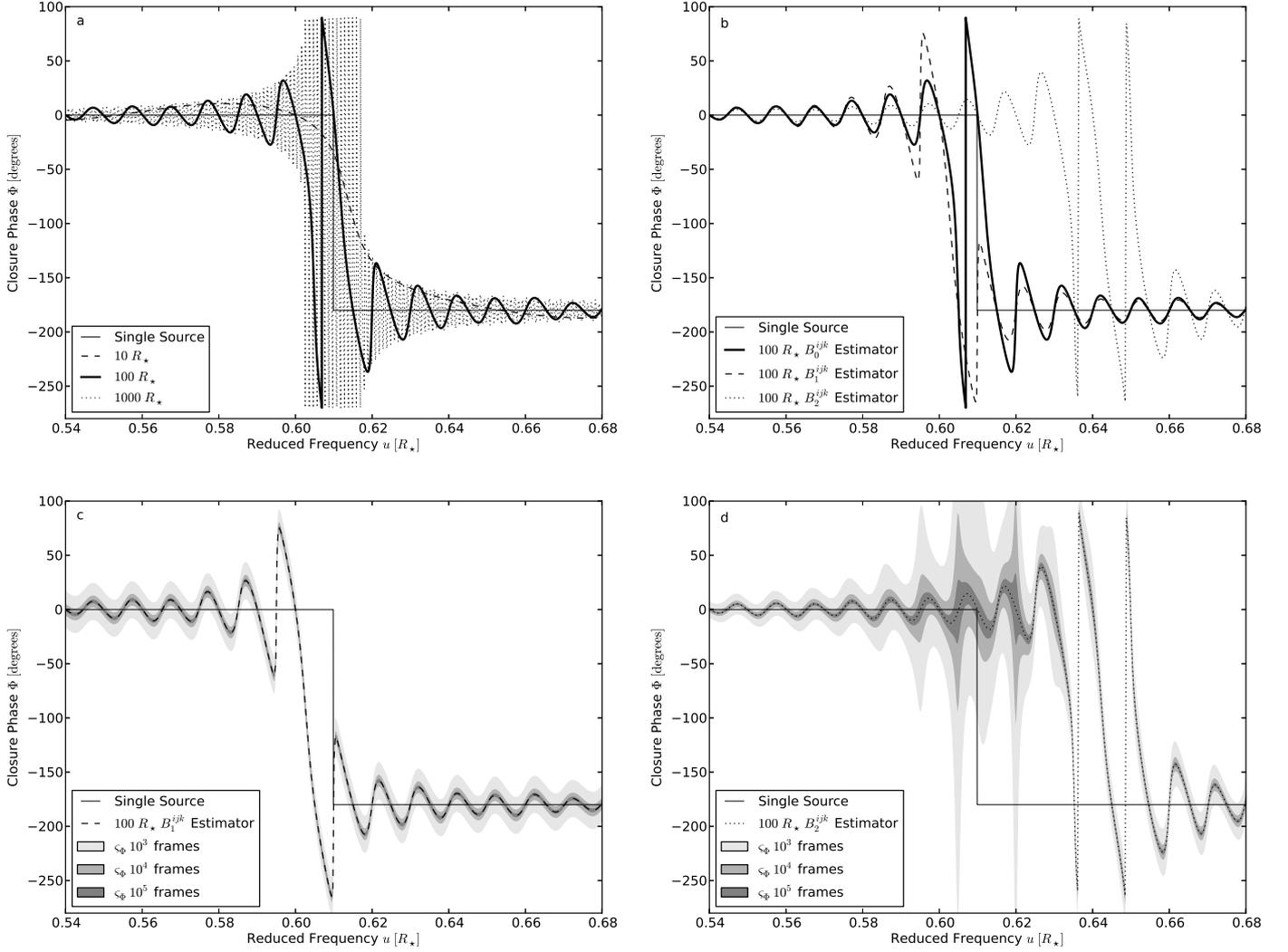}
\caption{Demonstration of the effect of bias on a closure phase measurement.
Plot \textbf{a}: The closure phase trace resulting from a binary pair
consisting of a resolved primary of radius $R_{\star}$ and an unresolved secondary with flux
ratio $r=0.01$ for separations $(s=10R_{\star}, 100R_{\star}, 1000R_{\star})$.
Plot \textbf{b}: The closure phase trace for the $100R_{\star}$ system recovered
by the three estimators. $B^{ijk}_{0}$ correctly returns the ideal
closure phase, whilst $B^{ijk}_{1}$ and $B^{ijk}_{2}$ do not. Both
plots show the single source closure phase signal (thin solid line) for
reference. Plots \textbf{c} and \textbf{d}
show the closure phase calculated using $B^{ijk}_{1}$ and $B^{ijk}_{2}$
respectively as in plot \textbf{b} along with the statistical uncertainty on
the measurement after $10^{3}$, $10^{4}$ and $10^{5}$ frames.}
	\label{fig:cpn}
\end{figure*}

The ideal bispectrum is extracted from
Eq.~(\ref{eqn:comp_vis_binary}) using Eq.~(\ref{eqn:B_first}) and
the closure phase is taken as the argument of the averaged bispectrum. The
closure phase signal around the first null in the visibility profile is shown in
Fig.~2 of \citet{chelli_phase_2009} and reproduced using our simulation model in
Fig.~\ref{fig:cpn}a in this paper for a number of companion separations
$(s=10R_{\star}, 100R_{\star}, 1000R_{\star})$. The closure phase is wrapped
modulo $2\pi$ and the frequencies $\{{u_{12}, u_{23}}, u_{13}\}$ are set to the
ratio $\{1:2:3\}$ on the detector and scaled by the radius of the primary,
$R_{\star}$. For simplicity, the baseline to fringe-frequency mapping is
chosen such that it preserves the ratios of the two. The specifics of the
mapping is not expected to have a significant impact on the bias. The closure
phase of the primary alone (single source) is shown for comparison.

%
We present in
Fig.~\ref{fig:cpn}b a simulation of the closure phase from a $100R_{\star}$
separation binary with flux ratio $0.01$. $B^{ijk}_{0}$ returns the
correct closure phase signal in agreement with the ideal case presented by
\citet{chelli_phase_2009}. $B^{ijk}_{1}$ and
$B^{ijk}_{2}$ are strongly affected by bias terms, meaning the
closure phase is significantly different. Depending which regions of the
closure phase trace are sampled by the interferometer, $B^{ijk}_{1}$
and $B^{ijk}_{2}$ would return an incorrect value for the primary
radius and companion flux ratio. Figs.~\ref{fig:cpn}c and \ref{fig:cpn}d
show the closure phase calculated using $B^{ijk}_{1}$ and $B^{ijk}_{2}$
respectively and the statistical uncertainty on the measurement after $10^{3}$,
$10^{4}$ and $10^{5}$ frames calculated using Eq.~(\ref{eq:cp_std_dev}). We see that even for
very short total integration times ($10^{3}$ frames is under a minute using
typical discrete integration times) the bias still produces a significant
contribution.

\section{Conclusions}

Interferometry offers exceptional angular resolution at the expense of
sensitivity, as the collecting area is, by necessity, smaller than that of a
single dish with the same resolution. In ground-based optical interferometry
these sensitivity limitations are accentuated by the need for short observations
to reduce the phase perturbations introduced by the atmosphere. The result is
that noise levels on individual observation frames are often high, exacerbated
by the need for fast read-out of the detector. Without understanding the
statistical properties of the noise, bias terms are
introduced when extracting information from the interferogram using the
non-linear power spectrum and bispectrum constructs. We have shown in this paper
that:

\begin{enumerate}
  \item Neglecting the cross terms that arise in the presence of both read
  noise and photon noise leads to bias on the extraction of interferometric
  observables resulting from the statistics of the noise.
  \item The statistical bias on the observables results in incorrect
  phisical parameters being extracted from observation data, for
  example the flux ratio of binary pairs or the diameter of a target as shown for the case of closure phase
  nulling.
  \end{enumerate}

To remedy this we have presented bias-free estimators for both the power
spectrum and bispectrum (and hence the closure phase) which can be used in the
presence of any combination of photon and read noise under any fringe encoding
scheme in which the measurement equation is linear. Our bias-free estimators are
presented in a format allowing easy implementation into current data reduction pipelines.


\begin{acknowledgements}
      We thank the anonymous referee for constructive comments and insiteful
      suggestions that have undoubtedly improved this paper.
\end{acknowledgements}

\bibliographystyle{bibtex/aa}
\bibliography{fringe_decomp}
 
 \Online

\begin{appendix} 


\section{Derivation of the bias-free bispectrum
estimator}\label{app:bispec_der}

The ideal, noise-free bispectrum is given by:
\begin{equation}
B^{ijk} = C^{ij}C^{jk}C^{ki} =
\sum_{p1}\sum_{p2}\sum_{p3} \Lambda_{p1}\Lambda_{p2}\Lambda_{p3}
H^{ij}_{p1}H^{jk}_{p2}H^{ki}_{p3},
\end{equation}
which can be split into:
\begin{eqnarray}
B^{ijk} &=& \sum_{p} \Lambda^{3}_{p}H^{ij}_{p}H^{jk}_{p}H^{ki}_{p} \nonumber\\
&+& \underset{\ p\ \neq\  p1}{\sum\sum}\Lambda_{p}^{2}\Lambda_{p1}H^{ij}_{p1}H^{jk}_{p}H^{ki}_{p}\nonumber\\
&+& \underset{\ p\ \neq\ 
p2}{\sum\sum}\Lambda_{p}^{2}\Lambda_{p2}H^{ij}_{p}H^{jk}_{p2}H^{ki}_{p}\nonumber\\ &+&
\underset{\ p\ \neq\  p3}{\sum\sum}\Lambda_{p}^{2}\Lambda_{p}H^{ij}_{p}H^{jk}_{p}H^{ki}_{p3}\nonumber\\
&+&\underset{\ p1\ \neq\ p2\ \neq\ p3}{\sum\sum\sum}\Lambda_{p1}
 \Lambda_{p2}\Lambda_{p3}
 H^{ij}_{p1}H^{jk}_{p2}H^{ki}_{p3}.\label{eqn:bispec_split}
\end{eqnarray}
The equivalent expression for real data embedded with noise when average over
frames is:
\begin{equation}
\left\langle c^{ij}c^{jk}c^{ki}\right\rangle=
 \sum_{p1}\sum_{p2}\sum_{p3}
 \left\langle i_{p1}i_{p2}i_{p3}\right\rangle
 H^{ij}_{p1}H^{jk}_{p2}H^{ki}_{p3},
\end{equation}
which splits into:
\begin{eqnarray}
\left\langle c^{ij}c^{jk}c^{ki}\right\rangle &=& 
\sum_{p}\left\langle i^{3}_{p}\right\rangle
 H^{ij}_{p}H^{jk}_{p}H^{ki}_{p} \nonumber\\
&+& \underset{\ p\ \neq\   p1}{\sum\sum}
  \left\langle i^{2}_{p}\right\rangle \left\langle i_{p1}\right\rangle
  H^{ij}_{p1}H^{jk}_{p}H^{ki}_{p}\nonumber\\
&+& \underset{\ p\ \neq\   p2}{\sum\sum}
  \left\langle i^{2}_{p}\right\rangle \left\langle
  i_{p2}\right\rangle H^{ij}_{p}H^{jk}_{p2}H^{ki}_{p}\nonumber\\ 
&+& \underset{\ p\ \neq\   p3}{\sum\sum}
  \left\langle i^{2}_{p}\right\rangle \left\langle
  i_{p3}\right\rangle H^{ij}_{p}H^{jk}_{p}H^{ki}_{p3}\nonumber\\
&+&\underset{\ p1\ \neq\ p2\ \neq\ p3}{\sum\sum\sum}
  \left\langle i_{p1}\right\rangle\left\langle i_{p2}\right\rangle
  \left\langle i_{p3}\right\rangle H^{ij}_{p1}H^{jk}_{p2}H^{ki}_{p3}.
\end{eqnarray}
Substitution of the noise model in
Eqs.~(\ref{eqn:noise_model_1}) to (\ref{eqn:noise_model_3}) gives the noisy
bispectrum in terms of the ideal interferogram and associated noise:
\begin{eqnarray}
\left\langle c^{ij}c^{jk}c^{ki}\right\rangle &=& 
\sum_{p}\left( \Lambda^{3}_{p} + 3\Lambda^{2}_{p} + \Lambda_{p} +
  	3\Lambda_{p}\sigma^{2}_{p} \right) H^{ij}_{p}H^{jk}_{p}H^{ki}_{p} \nonumber\\
&+& \underset{\ p\ \neq\  
  p1}{\sum\sum}\left( \Lambda^{2}_{p} + \Lambda_{p} + \sigma^{2}_{p}\right)\Lambda_{p1} 
  H^{ij}_{p1}H^{jk}_{p}H^{ki}_{p}\nonumber\\
&+& \underset{\ p\ \neq\  
  p2}{\sum\sum}\left( \Lambda^{2}_{p} + \Lambda_{p} + \sigma^{2}_{p}\right)\Lambda_{p2} 
  H^{ij}_{p}H^{jk}_{p2}H^{ki}_{p}\nonumber\\ 
&+& \underset{\ p\ \neq\  
  p3}{\sum\sum}\left( \Lambda^{2}_{p} + \Lambda_{p} + \sigma^{2}_{p}\right)\Lambda_{p3} 
  H^{ij}_{p}H^{jk}_{p}H^{ki}_{p3}\nonumber\\
&+&\underset{\ p1\ \neq\ p2\ \neq\ p3}{\sum\sum\sum}\Lambda_{p1}
 \Lambda_{p2}\Lambda_{p3} H^{ij}_{p1}H^{jk}_{p2}H^{ki}_{p3}.
\end{eqnarray}
Substituting the ideal bispectrum split sum from Eq.~(\ref{eqn:bispec_split})
gives:
\begin{eqnarray}
\left\langle c^{ij}c^{jk}c^{ki}\right\rangle &=& B^{ijk} + \sum_{p}\left(3\Lambda^{2}_{p} + \Lambda_{p} +
  	3\Lambda_{p}\sigma^{2}_{p} \right) H^{ij}_{p}H^{jk}_{p}H^{ki}_{p} \nonumber\\
&+& \underset{\ p\ \neq\  
  p1}{\sum\sum}\left( \Lambda_{p} + \sigma^{2}_{p}\right)\Lambda_{p1} 
  H^{ij}_{p1}H^{jk}_{p}H^{ki}_{p}\nonumber\\
&+& \underset{\ p\ \neq\  
  p2}{\sum\sum}\left(  \Lambda_{p} + \sigma^{2}_{p}\right)\Lambda_{p2} 
  H^{ij}_{p}H^{jk}_{p2}H^{ki}_{p}\nonumber\\ 
&+& \underset{\ p\ \neq\  
  p3}{\sum\sum}\left(  \Lambda_{p} + \sigma^{2}_{p}\right)\Lambda_{p3} 
  H^{ij}_{p}H^{jk}_{p}H^{ki}_{p3}.
\end{eqnarray}
By rearranging and substituting in for the ensemble average over the real
interferogram we find the ideal bispectrum in terms of the noisy interferogram:
\begin{eqnarray}
B^{ijk} &=& \left\langle
c^{ij}c^{jk}c^{ki}\right\rangle\nonumber\\
 &-& \left\langle\sum_{p}\left(3 i^{2}_{p} -2
 i_{p} + 3
 i_{p}\sigma^{2}_{p} - 3
 \sigma^{2}_{p} \right) H^{ij}_{p}H^{jk}_{p}H^{ki}_{p}\right\rangle
 \nonumber\\ &-& \underset{\ p\ \neq\ p1}{\sum\sum} \left\langle i_{p} +
 \sigma^{2}_{p}\right\rangle\left\langle i_{p1}\right\rangle
 H^{ij}_{p1}H^{jk}_{p}H^{ki}_{p}\nonumber\\ 
 &-& \underset{\ p\ \neq\ p2}{\sum\sum} \left\langle i_{p} +
 \sigma^{2}_{p}\right\rangle\left\langle i_{p2}\right\rangle
 H^{ij}_{p}H^{jk}_{p2}H^{ki}_{p}\nonumber\\ 
&-& \underset{\ p\ \neq\ p3}{\sum\sum} \left\langle i_{p} +
 \sigma^{2}_{p}\right\rangle\left\langle i_{p3}\right\rangle
 H^{ij}_{p}H^{jk}_{p}H^{ki}_{p3}.
\end{eqnarray}
Expanding out the double summations using the fact that  $\underset{\ a\ \neq\
b}{\sum\sum} = \underset{\ a\ \ \
b}{\sum\sum} - \underset{a=b}{\sum}$ gives:
\begin{eqnarray}
B^{ijk}&=& \left\langle
c^{ij}c^{jk}c^{ki}\right\rangle\nonumber\\
 &-& \left\langle\sum_{p}\left(3i^{2}_{p} -2
 i_{p} + 3
 i_{p}\sigma^{2}_{p} - 3
 \sigma^{2}_{p} \right)
 H^{ij}_{p}H^{jk}_{p}H^{ki}_{p}\right\rangle \nonumber\\
 &+& 3\left\langle\sum_{p}\left(i^{2}_{p} +
 i_{p}\sigma^{2}_{p}\right)
 H^{ij}_{p}H^{jk}_{p}H^{ki}_{p}\right\rangle \nonumber\\
  &-& \left\langle\sum_{p}\left(i_{p} +
 \sigma^{2}_{p}\right) H^{jk}_{p}H^{ki}_{p}\sum_{p1} i_{p1}
 H^{ij}_{p1}\right\rangle\nonumber\\
 &-&\left\langle\sum_{p}\left(i_{p} +
 \sigma^{2}_{p}\right) H^{ij}_{p}H^{ki}_{p}\sum_{p2} i_{p2}
 H^{jk}_{p2}\right\rangle\nonumber\\ 
 &-&\left\langle\sum_{p}\left(i_{p} +
 \sigma^{2}_{p}\right) H^{ij}_{p}H^{jk}_{p}\sum_{p3} i_{p3}
 H^{ki}_{p3}\right\rangle.
\end{eqnarray}
Cancelling terms and noting that, $\sum_{p}i_{p}H^{ij}_{p} = c^{ij}$,
we find the bias free bispectrum estimator:
\begin{eqnarray}
B^{ijk}_{\mathrm{bias-free}}&=& 
c^{ij}c^{jk}c^{ki}\nonumber\\
 &-&  c^{ij}\sum_{p}\left(i_{p} +
 \sigma^{2}_{p}\right) H^{jk}_{p}H^{ki}_{p} \nonumber\\
 &-&c^{jk}\sum_{p}\left(i_{p} +
 \sigma^{2}_{p}\right) H^{ij}_{p}H^{ki}_{p} \nonumber\\ 
 &-& c^{ki}\sum_{p}\left(i_{p} +
 \sigma^{2}_{p}\right) H^{ij}_{p}H^{jk}_{p} \nonumber\\
 &+& \sum_{p}\left(2 i_{p}
 +3\sigma^{2}_{p}\right)H^{ij}_{p}H^{jk}_{p}H^{ki}_{p} 
\end{eqnarray}

\end{appendix}

\end{document}